\documentstyle{article}

\begin{document}
\begin{center}
{\Large\bf Electron-positron pair production \\ in the Aharonov-Bohm 
potential}\\[1cm]

{\large
Vladimir D.~Skarzhinsky\footnote{e-mail:
vdskarzh@sgi.lpi.msk.su}}\\
\medskip
Fakult\"at f\"ur Physik der Universit\"at Konstanz,
Postfach 5560, D 78434 Konstanz, Germany\\
and P.~N.~Lebedev Physical Institute,
Leninsky prospect 53, Moscow 117924, Russia\\[0.5cm]
{\large J\"urgen Audretsch\footnote{e-mail:
Juergen.Audretsch@uni-konstanz.de}
and Ulf Jasper\footnote{e-mail:
Ulf.Jasper@uni-konstanz.de}
}\\[0.5cm]
Fakult\"at f\"ur Physik der Universit\"at Konstanz,
Postfach 5560, D 78434 Konstanz, Germany\\
\bigskip
(Phys.Rev.,D, {\bf 53}, 2190, 1996)
\end{center}

\begin{abstract}
In the framework of QED we evaluate the cross section for
electron-positron pair production by a single photon in the presence
of the external Aharonov-Bohm potential in first order of perturbation
theory. We analyse energy, angular
and polarization distributions at different energy regimes: near the
threshold and at high photon energies.

\medskip

\noindent PACS numbers: 03.65.Bz, 12.20.-m
\end{abstract}

\vspace{1cm}


\section{Introduction}

In a previous paper \cite{Audretsch95} we investigated the bremsstrahlung 
process for relativistic electrons scattered by the external Aharonov-Bohm 
(AB) potential (magnetic string). This process is supposed to be the most 
significant one among those accompanying the AB scattering \cite{Aharonov59}.
The AB effect -- the influence of magnetic fluxes on quantum systems -- can be 
adequately interpreted by means of {\it phase factors} \cite{Yang75}
$\exp\left(ie\oint A_{\mu}dx_{\mu}\right)$ which produce phase shifts in wave 
functions of charged particles. A number of remarkable experiments was made to 
observe the resulting interference pattern of an electron beam scattered by a 
thin solenoid. For a comprehensive review see \cite{Olariu85, Peshkin89}. 
In solid state physics the manifestation of the AB effect brought new 
unexpected results \cite{Lee85, Bergmann84}.

In addition to the bremsstrahlung process there exist other important quantum effects in the presence of the external AB field. We consider here in the framework of QED the production of an electron-positron pair by a single photon in first order. This process, as other analogous quantum processes, is possible only in the presence of external fields which provide the necessary momentum transfer. It happens, for example, in the Coulomb field \cite{Akhiezer65} or a uniform magnetic field \cite{Klepikov54} - \cite{Zharkov65}. In these cases there are external local forces which influence the motion of the created charged particles. In the AB case, however, the pair creation seems to be somehow mysterious since it happens due to a global, topological reason. In fact the AB field provides the violation of the momentum conservation law. Therefore the mechanism that permits pair production bears some resemblance with processes near cosmic strings \cite{Skarzhinsky94,Audretsch94}.

The theoretical study of the AB scattering for the Dirac electron 
\cite{Hagen90,Hagen91} raised a mathematical problem related with the 
correct description of the behavior of electron wave functions on the 
magnetic string. We do not discuss this problem here but refer to 
\cite{Hagen90,Hagen91,Audretsch95}. 
The issue of spin changes slightly the interpretation of the AB effect. 
The interaction between spin and magnetic field leads to wave functions of
Dirac particles which do not vanish on the magnetic string and thus, in a way, 
a local element is added to the non-locality of the AB effect.

\medskip

The paper is organized as follows. In section 2 we consider briefly the 
Dirac equation in the presence of the AB potential and work out the 
electron and positron 
wave functions characterized by quantum numbers of a complete set of commuting 
operators. The exact scattering wave function for electrons and positrons are
expressed in terms of partial waves. In section 3 the matrix element for the 
pair production by a single photon is calculated and the effective differential 
cross section is evaluated. We analyze the behavior of the differential and 
total cross section at different energy regimes and discuss their particular
features for the Dirac electron in sections 4 and 5.

\medskip

We use units such that $\hbar=c=1$ and take $e<0$ for the electron charge.


\section{The electron and positron solutions to the Dirac equation in
the Aharonov-Bohm potential}

The Dirac equation in an external magnetic field reads 
\begin{equation} \label{de}
i\partial_{t}\psi = H\psi,\quad H = \alpha_{i}(p_{i} - eA_{i}) +
\beta M
\end{equation}
where $e$ is the electron charge. For matrices $\alpha$ and $\beta$ we use
\begin{equation}
\alpha_{i} = \pmatrix{      0    & \sigma_{i} \cr
                      \sigma_{i} &       0 \cr}, \quad
\beta = \pmatrix{      1   &  0\cr
                       0   &  -1 \cr}.
\end{equation}
In cylindrical coordinates $(\rho,\;\varphi,\;z)$ the kinetic momenta are 
given by
\begin{equation}
\pi_{\rho} = p_{\rho} = - i\partial_{\rho},\quad \pi_{\varphi} =
p_{\varphi} - eA_{\varphi}=-{i \over
\rho}\partial_{\varphi}-eA_{\varphi},\quad p_{3} = -i \partial_{z}
\end{equation}
where
\begin{equation}
\sigma_{\rho} = \sigma_1 \cos{\varphi} + \sigma_2 \sin{\varphi},
\quad \sigma_{\varphi} = -\sigma_1 \sin{\varphi} + \sigma_2
\cos{\varphi}
\end{equation}
and $\sigma_i$ are the Pauli matrices.

The vector potential for the {\it pure AB case} (magnetic string) has a 
nonzero angular compo\-nent \cite{Aharonov59}
\begin{equation}\label{abp}
eA_{\varphi} = {e\Phi \over 2\pi\rho} = -{\Phi \over \Phi_{0}\rho} =
{\phi\over\rho},
\end{equation}
where $\Phi$ is the magnetic flux and $\Phi_{0}=2\pi \hbar c/|e|$ is the
magnetic flux quantum. It corresponds to a magnetic field with support on the 
$z$-axis
\begin{equation}
B_z = {2\phi\over e\rho}\delta (\rho)
\end{equation}
which points to the positive (negative) $z$ direction for $\phi<0 \; (\phi>0)$. 
Note that it is the fractional part $\delta$ of the magnetic flux $\phi =
N+\delta,\; 0<\delta<1$ which produces all physical effects. Its
integral part $N$ appears as a phase factor $\exp(iN\varphi)$ in
solutions of the Dirac equation.

The exact solution of the Dirac equation for the scattering problem
in the external AB field can be written in an integral
form as it was done for the Schr\"odinger equation in the original
paper by Aharonov and Bohm \cite{Aharonov59}. 
For our problem, however, cylindrical modes are more convenient.

For the Dirac equation in the AB field the complete set of commuting operators 
is
\begin{eqnarray}
\hat{H}\,,\quad
\hat{p_3} := -i\partial_z\,,\quad
\hat{J_3} := -i\partial_{\varphi}+{1 \over 2} \Sigma_3 \,,\quad 
\label{s3}
\hat{S_3} := \beta\Sigma_3+\gamma \; {p_3\over M}\,
\end{eqnarray}
where $\gamma := \pmatrix{0 & 1 \cr
                   1 & 0 \cr} $.
The corresponding eigenvalue equations are given by
\begin{eqnarray} \label{ev}
\hat{H}\psi &=& E\psi\,,\\
\hat{p_3}\psi &=& p_3\psi\,,\\
\hat{J_3}\psi &=& j_3 \psi\,,\\
\hat{S_3}\psi &=& s\psi\,,
\end{eqnarray}
where $E = \sqrt{p_\perp^2 + p_3^2 +M^2}$ is the energy, $p_3$ and $j_3$
are the $z$-components of linear and total angular momentum respectively;
$p_\perp$ denotes the radial momentum.
The eigenvalue of $\hat{J}_3$ is half-integer and we rewrite it by introducing
$l$, $j_3 =: l + N +1/2$. Here $l$ is an integer number and N is
fixed as above.
Note that $l+N$ denotes an integral part of the eigenvalue of $\hat{J_3}$ in 
contrast to the usual convention. The corresponding separation of a factor 
$\exp(i N \varphi)$ in the solutions of the Dirac equation will turn out to 
be convenient in the following calculations.
We introduced in eq.~(\ref{s3}) the operator $\hat{S}_3$ and not the helicity 
operator $\hat{S}_t = \Sigma_i (p_i -e A_i) / p$ which is often used.
Both of these operators commute in 
the relativistic case with the operators $\hat{H}$, $\hat{p}_3$ and 
$\hat{J}_3$, when a magnetic field of a fixed direction is present. 
This can be seen for example in \cite{Sokolov68}. We prefer to use $\hat{S}_3$ 
because in the nonrelativistic limit, which will be treated below, it describes 
the spin projection along the direction of the magnetic field.
Its eigenvalue is given by $s=\pm \sqrt{1+p_3^2/ M^2}$.
Solving these eigenvalue equations leads to a radial solution of Bessel type.

As independent radial solutions we choose Bessel functions of the first kind 
of positive and negative orders. Then the normalization condition for the 
partial modes with quantum numbers $j = (p_{\perp},p_3,l,s)$
\begin{equation} \label{nc}
\int d^3 x \psi^{\dagger}(j,x) \psi(j^{\prime},x) = \delta_{j, j^{
\prime}} = {\delta_{s, s^{ \prime}} \delta_{l, l^{ \prime}}}{
\delta(p_{3} - p_{3^{ \prime}})} {\delta(p_{ \perp} - p_{ \perp}^{
\prime}) \over  \sqrt{p_{\perp} p_{ \perp}^{ \prime}}},
\end{equation}
fixes the solutions (for electron states with $E>0$) of these equations
for values of $l$ outside of the interval $-1 < l-\delta <0$ thus removing 
Bessel functions with negative order which are not square integrable. One 
finds, for $l\neq 0$, 
\begin{equation}\label{es}
\psi_{e}(j,x) = {1\over 2\pi}{1\over\sqrt{2E_p}}\, e^{-iE_p t +
ip_3z}\, e^{iN\varphi}\, e^{i{\pi\over 2}|l|}\,\pmatrix{ u \cr
v\cr},
\end{equation}
where
\begin{eqnarray}\label{uv1}
u &=& {1\over \sqrt{2s}}\pmatrix{\sqrt{E_p + sM}\sqrt{s+1}\;
J_{\nu_1}(p_{\perp}\rho)\, e^{il\varphi}\cr
i\epsilon_3\epsilon_{l}\sqrt{E_p - sM}\sqrt{s-1}\;
J_{\nu_2}(p_{\perp}\rho)\, e^{i(l +1)\varphi} \cr}, \\
\nonumber\\
\label{uv2}
v &=& {1\over \sqrt{2s}}\pmatrix{\epsilon_3 \sqrt{E_p + sM}\sqrt{s-1}
\;J_{\nu_1}(p_{\perp}\rho)\, e^{il\varphi}\cr
i\epsilon_{l}\sqrt{E_p - sM}\sqrt{s+1} \;
J_{\nu_2}(p_{\perp}\rho)\, e^{i(l +1)\varphi} \cr}
\end{eqnarray}
and
\begin{equation}
p_{\perp} := \sqrt{p^2 - p^2_3} = \sqrt{E_p^2 - M^2 - p^2_3},\quad s =
\pm \sqrt{1+{p_3^2\over M^2}},\quad \epsilon_3 := {\rm sign}(s p_3),
\end{equation}
\begin{equation} \label{nu}
\nu_1 := \cases{\;\; l-\delta & \cr -l+\delta & \cr}, \quad
\nu_2 := \cases{\;\;l+1-\delta & \cr -l-1+\delta & \cr}, \quad
\epsilon_l := \cases{\;\;1 &\quad if $l \geq 0$\cr -1 &\quad if $l<0.
$\cr}
\end{equation}
Since the normalization condition method does not apply for $l=0$ this case 
needs a separate discussion, which has been done in \cite{pragm}. 
Fortunately it turns out that the corresponding solution is of the same form as for $l\neq0$. Therefore the expressions above are valid not only for $l\neq0$ but also for $l=0$ so that it is allowed to include this case in (\ref{nu}).
The critical mode $l=0$
contains the irregular but square integrable Bessel functions of orders 
$-\delta$ and $-1+\delta.$ Their inevitable appearance in solutions of the 
Dirac equation is an obvious consequence of the interaction between
spin and magnetic field. This problem is a part of a general problem of the 
self-adjoint extension for the Hamilton operator in the presence of a singular 
potential ({\it pure AB case}), and it was discussed in the previous paper 
\cite{Audretsch95} (see also \cite{Hagen90, Coutinho94}). In \cite{pragm} we 
presented an alternative method of treating the self-adjointness problem.

The complete set of solutions of the Dirac equations includes the
negative energy electron states. Instead of them we introduce positron
states $\psi_{p}$ with $E>0$ which can be obtained from electron
states of negative energy by the charge conjugation operation
\begin{equation} \label{conj}
\psi \rightarrow \psi_{c} = C\bar \psi_{transp}\,,\;\; C =
\alpha_{2}
\end{equation}
and replacing $e$ by $-e$.
$\psi_{c}$ obeys the free Dirac equation as well as $\psi$ does but with
opposite sign of the electric charge and has quantum
numbers $E, -p_3, -j_3, s.$ One needs to replace $p_3 \rightarrow
-p_3,\;j_3 \rightarrow -j_3 \;(l \rightarrow -l-1)$ in
the electron state of negative energy to obtain positron state with
quantum numbers $E, p_3, j_3, s$. 

The electron-positron field operator reads
\begin{equation} \label{epfo1}
\psi(x,t) = \int d\mu_j [\psi_{e}(j,x) a_j + \psi^{c}_{p}(j,x)
b^{\dagger}_{j}],
\end{equation}
with $a_j$ and $b_j$ being the annihilation operators for the electron and
positron with given quantum numbers. It contains positive frequency
functions $\psi_e$ (electron states) and negative frequency
functions $\psi^c_p$ (positron states),
\begin{equation}\label{ps}
\psi^{c}_{p}(j,x) = {1\over
2\pi}{1\over\sqrt{2E_p}}\, e^{iEt-ip_3z}\, e^{iN\varphi} \, e^{i{\pi\over
2}|l|}
\pmatrix{ y \cr
w\cr},
\end{equation}
where
\begin{eqnarray} \label{yw1}
y &=& {1\over\sqrt{2s}}
\pmatrix{i\epsilon_l \sqrt{E_p-sM}\sqrt{s+1}\;
J_{\nu'_2}(p_{\perp}\rho)\, e^{-i(l+1)\varphi}\cr
\epsilon_3 \sqrt{E_p+sM}\sqrt{s-1}\;J_{\nu'_1}(p_{\perp}\rho)\, e^{-il
\varphi} \cr}, \\
\nonumber\\
\label{yw2}
w &=& - {1\over\sqrt{2s}}
\pmatrix{i\epsilon_3 \epsilon_l\sqrt{E_p-sM}\sqrt{s-1}\;
J_{\nu'_2}(p_{\perp}\rho)\, e^{-i(l+1)\varphi} \cr
\sqrt{E_p+sM}\sqrt{s+1}\;J_{\nu'_1}(p_{\perp}\rho)\, e^{-il\varphi} \cr}
\end{eqnarray}
with
\begin{equation} \label{nu2}
\nu'_1 := \cases{\;\; l+\delta & \cr -l-\delta & \cr}, \quad
\nu'_2 := \cases{\;\;l+1+\delta & \quad if $l \geq 0$\cr -l-1-\delta &
\quad if $l < 0$ \cr}.
\end{equation}

The expressions (\ref{es})--(\ref{nu}) and (\ref{ps})--(\ref{nu2})
present the partial electron and positron wave functions in terms of 
cylindrical modes. These states do not describe outgoing particles with 
definite linear momenta at infinity. In order to calculate the cross section 
of the pair production process we need the {\em electron and positron 
scattering wave functions}. In external fields there exist two independent 
exact solutions of the Dirac equation which behave at large distances like a 
plane wave (propagating in the direction $\vec{p}$ given by $p_x = p_{\perp} 
\cos\varphi_p, \; p_y = p_{\perp} \sin\varphi_p, \; p_z$) plus an outgoing or 
ingoing cylindrical waves, correspondingly. For outgoing particles we need to 
take wave functions which contain ingoing cylindrical waves. In this case the 
interaction of the created electron and positron with the external magnetic 
field will be described correctly \cite{Akhiezer65}.

The corresponding scattering wave functions can be obtained by superpositions 
of the cylindrical modes.
\begin{equation} \label{epw}
\Psi_{e} (J, x) := \sum_l c_l^{(e)}\;\psi_{e} (j_p, x)
\end{equation}
and
\begin{equation} \label{ppw}
\Psi_{p}^c (J ,t) := \sum_n c_n^{(p)}\;\psi_{p}^c (j_q, x)
\end{equation}
with the coefficients
\begin{equation} \label{cc}
c_l^{(e)} := e^{-il\varphi_p}\;e^{-i{\pi\over 2}\epsilon_l \delta}\,, \quad
c_n^{(p)} := e^{i(n+1)(\varphi_q+\pi)}\;e^{i{\pi\over 2}\epsilon_n
\delta}
\end{equation}
where $J$ is a collective index for the linear momentum at infinity and $s$.

In the terms of the wave functions (\ref{epw}) and (\ref{ppw}) the 
electron-positron field operator reads
\begin{equation} \label{epfo2}
\psi(x,t) = \int d\mu_J [\Psi_{e}(J,x) a_J + \Psi^{c}_{p}(J,x)
b^{\dagger}_{J}],
\end{equation}
with $a_J$ and $b_J$ being the annihilation operators for the electron and
positron with quantum numbers of the scattering states. 

The external AB field has no influence on the {\it photon
wave function}. In cylindrical coordinates it reads
\begin{equation}\label{vp}
A_{\mu}^{\lambda}(\vec{k}, x) =
{e_{\mu}^{(\lambda)}\over\sqrt{2\omega_k}}e^{-i\omega_k t + ik_3 z}
e^{ik_{\perp}\rho \cos(\varphi-\varphi_k)}
\end{equation}
where the polarization vectors 
\begin{equation}\label{polv}
e^{(\sigma)} := (0,\; -\sin \varphi_k, \;\cos \varphi_k, \; 0), \quad
e^{(\pi)} := {1\over \omega_k}(0,\; -k_3\cos \varphi_k,\; -k_3\sin
\varphi_k,\; k_{\perp})
\end{equation}
correspond to two linear transversal polarization states. In the coordinate 
frame with $k_3=0$ in which we will perform all calculations, the polarization 
vector $e^{(\pi)}$ is directed along $z$-axis and $e^{(\sigma)}$ is orthogonal 
to the magnetic string.


\section{Matrix elements and differential cross sections for pair
production  by a single photon $\gamma \rightarrow e^{-} + e^{+}$}

The differential cross section of the pair production process describes the 
distribution of the created particles with respect to their quantum numbers. 
For these one may take the angular momenta which correspond to cylindrical 
modes. But usually final states are related to plane wave states, and in our 
case to the scattering states (\ref{epw}), (\ref{ppw}). The cylindrical modes 
have a vanishing radial flux and therefore do not describe ingoing or outgoing 
particles. They are, however, convenient for calculating matrix elements, and 
we use these matrix elements as starting point for calculation of the 
differential cross section which refers to scattering states. As far as the 
total cross section is concerned one can use any final states.

\subsection{Matrix elements for cylindrical modes}

The matrix element for pair production of an electron with quantum
numbers $j_p = (p_{\perp}, p_3,l,s)$ and a positron with quantum
numbers $j_q = (q_{\perp}, q_3,n,r)$ by a single photon with quantum
numbers ($\vec{k},\; \lambda)$ for physical states $\lambda =
\sigma,\; \pi$ has the usual form
\begin{equation} \label{me1}
\widetilde{M}(j_p, j_q; \vec{k}, \lambda) 
= -i \langle j_q, j_p|S^{(1)}|\vec{k}, \lambda \rangle
= -e \int {d^4x}\bar\psi_{e}(j_{p},x)
\;A^{\lambda}_{\mu}(\vec{k},x)\gamma_{\mu}\; \psi_{p}^c (j_{q},x)
\end{equation}
whereby gamma matrices are written in terms of Pauli matrices as
\begin{equation} \label{gamma}
\gamma_i = \pmatrix{
         0    & \sigma_i \cr
    - \sigma_i &       0 \cr},
\end{equation}
so that
\begin{equation} \label{alphas}
e_{\mu}^{(\lambda)} \gamma_{\mu} = \pmatrix{
        0    &  \alpha_{\lambda} \cr
      -\alpha_{\lambda}  &  0  \cr}, \;
\alpha_{\sigma} = \pmatrix{
        0    & -i e^{-i\varphi_k} \cr
        i e^{i\varphi_k}&  0  \cr}, \;
\alpha_{\pi} = \pmatrix{
       k_{\perp}   & -k_3 e^{-i\varphi_k} \cr
     -k_3 e^{i\varphi_k} &  - k_{\perp} \cr}{1\over\omega_k}.
\end{equation}

Using the expressions (\ref{es}), (\ref{ps}) and (\ref{vp}), we can rewrite 
the matrix element (\ref{me1}) in the form
\begin{equation} \label{me2}
\widetilde{M}_{\lambda}(j_p, j_q) =-e{1\over{2\sqrt{2\omega_{k}E_{q}E_{p}}}}
e^{-i{\pi\over
2}(|l|-|n|)}\delta(E_p+E_q-\omega_{k})\delta(p_3+q_3-k_3)
\;m_{\lambda}
\end{equation}
with
\begin{eqnarray} \label{me3}
m_{\lambda} :&=& \int\rho d\rho d\varphi
e^{ik_{\perp}\rho\cos(\varphi-\varphi_k)}
\;\left[u^{\dagger}(p)\alpha_{\lambda}w(q) +
v^{\dagger}(p)\alpha_{\lambda}y(q)\right ] \cr
&=&e^{-i(l+n+1)\varphi_k}\;\int\rho d\rho d\varphi
e^{ik_{\perp}\rho\cos(\varphi-\varphi_k)} K_{\lambda}(\rho, \varphi).
\end{eqnarray}

The Dirac equation (\ref{de}) in the external AB field is
invariant under boost transformations along the string direction.
This means that it is sufficient to treat the case of normal
incidence of the photon on the magnetic string, and therefore we may perform 
all calculations in the coordinate system in which $k_3=0$. No information 
will be lost but calculations become simpler in this case.

For the polarization state $\lambda = \sigma$ we have
\begin{eqnarray} \label{Ks}
K_{\sigma}(\rho, \varphi) &:=& 
i R_{\sigma}\left[\sqrt{E_p+sM}\sqrt{E_q+rM}\; J_{\nu_1}(p_{\perp}\rho)
J_{\nu'_1}(q_{\perp}\rho) \; e^{-i(l+n)(\varphi-\varphi_k)}\right. \nonumber\\
&& + \left.\epsilon_l\epsilon_n \sqrt{E_p-sM}\sqrt{E_q-rM}\;
J_{\nu_2}(p_{\perp}\rho) J_{\nu'_2}(q_{\perp}\rho)\;
e^{-i(l+n+2)(\varphi-\varphi_k)}\right],
\end{eqnarray}
with
\begin{equation}
R_{\sigma} := {1\over 2\sqrt{sr}} \left[\sqrt{s+1}\sqrt{r+1}
-\epsilon_3(p)\epsilon_3(q)\sqrt{s-1}\sqrt{r-1}\;\right]
\end{equation}
and for the polarization state $\lambda = \pi$
\begin{eqnarray} \label{Kp}
K_{\pi}(\rho, \varphi) &:=& 
i R_{\pi}\left[\epsilon_n
\sqrt{E_p+sM}\sqrt{E_q-rM}\; J_{\nu_1}(p_{\perp}\rho)
J_{\nu'_2}(q_{\perp}\rho)  \right. \nonumber\\
&&\left. - \epsilon_l \sqrt{E_p-sM}\sqrt{E_q+rM}
\;J_{\nu_2}(p_{\perp}\rho) J_{\nu'_1}(q_{\perp}\rho)
\right]e^{-i(l+n+1)(\varphi-\varphi_k)}
\end{eqnarray}
with
\begin{equation}
R_{\pi} := {1\over 2\sqrt{sr}} \left[\epsilon_3(p)\sqrt{s-1}\sqrt{r+1}
-\epsilon_3(q)\sqrt{s+1}\sqrt{r-1}\;\right].
\end{equation}

Integrating over $\varphi$ we obtain
\begin{eqnarray} \label{ms1}
m_{\sigma} &=& 2\pi i e^{-i(l+n+1)\varphi_k} e^{i{\pi\over 2}(l+n)}
R_{\sigma} \nonumber\\
&&
\times \int\rho d\rho \left[\sqrt{E_p+sM}\sqrt{E_q+rM}\;
J_{\nu_1}(p_{\perp}\rho)
J_{\nu'_1}(q_{\perp}\rho) J_{l+n}(k_{\perp}\rho) \right.\nonumber\\
&&
- \left. \epsilon_l \epsilon_n \sqrt{E_p-sM}\sqrt{E_q-rM}\;
J_{\nu_2}(p_{\perp}\rho) J_{\nu'_2}(q_{\perp}\rho)
J_{l+n+2}(k_{\perp}\rho)\right]
\end{eqnarray}
and
\begin{eqnarray} \label{mp1}
m_{\pi} &=& 2\pi i e^{-i(l+n+1)\varphi_k} e^{i{\pi\over 2}(l+n+1)}
R_{\pi} \nonumber\\
&&
\times \int\rho d\rho \left[\epsilon_n \sqrt{E_p+sM}\sqrt{E_q-rM}\;
J_{\nu_1}(p_{\perp}\rho) J_{\nu'_2}(q_{\perp}\rho)
J_{l+n+1}(k_{\perp}\rho) \right. \nonumber\\
&& - \left.
\epsilon_l \sqrt{E_p-sM}\sqrt{E_q+rM}\; J_{\nu_2}(p_{\perp}\rho)
J_{\nu'_1}(q_{\perp}\rho) J_{l+n+1}(k_{\perp}\rho) \right].
\end{eqnarray}

It follows from the energy conservation law, $\omega_k=E_p+E_q,$ that the
photon's radial momentum obeys the inequality $k_{\perp} > p_{\perp} 
+q_{\perp}$. The excess of radial momentum, $k_\perp - (p_\perp + q_\perp)$, 
is transmitted to the flux tube. For this case, using formulae 
[6.578(3), 6.522(14)] of \cite{Gradshteyn80}, one can see that the integrals 
over $\rho$ vanish unless $(l+{1\over 2})(n+{1\over 2}) < 0$. This inequality 
is fulfilled at $l\geq 0, n<0$ and $n\geq 0, l<0$, and the nonvanishing 
integrals are of the type
\begin{eqnarray} \label{int}
J(\alpha, \beta) &:=& \int_{0}^{ \infty} \rho d\rho
J_{\alpha}(p_{\perp}\rho\sin{A}\cos{B})
J_{\beta}(q_{\perp}\rho\cos{A}\sin{B})
J_{\beta-\alpha}(k_{\perp}\rho) \nonumber\\
&=& {2\sin\pi\alpha\over \pi k_{\perp}^2
\cos(A+B)\cos(A-B)}\left({\sin{A} \over \cos{B}}
\right)^{\alpha}\left({\sin{B} \over \cos{A}}
\right)^{\beta}
\end{eqnarray}
with $ p_{\perp} = k_{\perp}\sin{A}\cos{B}, \; q_{\perp} =
k_{\perp}\sin{B}\cos{A}.$

Denoting
\begin{equation}  \label{ab}
a := {\sin A\over\cos B}={p_{\perp}\over E_p+\sqrt{q^2_3+M^2}},\quad
b := {\sin B\over\cos A}={q_{\perp}\over E_q+\sqrt{q^2_3+M^2}}
\end{equation}
we have in terms of the integral (\ref{int}) for the matrix elements
(\ref{ms1})
\begin{eqnarray} \label{ms2}
m_{\sigma} &=& 2\pi i e^{-i(l+n+1)\varphi_k} e^{i{\pi\over 2}(l+n)}
R_{\sigma} \nonumber\\
&&
\times\left\{\Theta(l\geq 0)\Theta(n<0)(-1)^{l+n}
\left[\sqrt{E_p+sM}\sqrt{E_q+rM} \; J(l-\delta, -n-\delta)
\right.\right. \nonumber\\
&& +\left.\left.\sqrt{E_p-sM}\sqrt{E_q-rM}\;J(l+1-\delta,
-n-1-\delta) \right] \right. \nonumber\\
&& +\left.\Theta(l<0)\Theta(n\geq 0)\left[\sqrt{E_p+sM}\sqrt{E_q+rM}\;
J(-l+\delta, n+\delta) \right.\right. \nonumber\\
&& + \left.\left. \sqrt{E_p-sM}\sqrt{E_q-rM}\;J(-l-1+\delta,
n+1+\delta) \right]\right\}   \nonumber\\
\cr
&=& - {4i R_{\sigma}\; e^{-i(l+n+1)\varphi_k +i{\pi\over 2}|l-n|}\;
\sin\pi\delta\over
\sqrt{k_{\perp}^4-2k_{\perp}^2(p_{\perp}^2+q_{\perp}^2)
+(p_{\perp}^2-q_{\perp}^2)^2}}\;a^{|l|}\;b^{|n|} \nonumber\\
&& \times \left[\Theta(l\geq 0)\Theta(n<0)(ab)^{-\delta}
\left(\sqrt{E_p+sM}\sqrt{E_q+rM} - {a\over
b}\sqrt{E_p-sM}\sqrt{E_q-rM} \right)\right. \nonumber\\
&& - \left. \Theta(l<0)\Theta(n\geq 0) (ab)^{\delta}
\left(\sqrt{E_p+sM}\sqrt{E_q+rM} - {b\over
a}\sqrt{E_p-sM}\sqrt{E_q-rM} \right)\right]
\end{eqnarray}
and for the matrix element (\ref{mp1})
\begin{eqnarray} \label{mp2}
m_{\pi} &=& - 2\pi i e^{-i(l+n+1)\varphi_k} e^{i{\pi\over 2}(l+n+1)}
R_{\pi}  \nonumber\\
&& \times \left\{\Theta(l\geq 0)\Theta(n<0)(-1)^{l+n+1}
\left[\sqrt{E_p+sM}\sqrt{E_q-rM} \; J(l-\delta, -n-1-\delta)
\right.\right. \nonumber\\
&& +\left.\left.\sqrt{E_p-sM}\sqrt{E_q+rM}\;J(l+1-\delta, -n-\delta)
\right] \right.  \nonumber\\
&& - \left.\Theta(l<0)\Theta(n\geq 0)\left[\sqrt{E_p+sM}\sqrt{E_q-rM}\;
J(-l+\delta, n+1+\delta) \right.\right. \nonumber\\
&& + \left.\left. \sqrt{E_p-sM}\sqrt{E_q+rM}\;J(-l-1+\delta, n+\delta)
\right]\right\} \nonumber\\
\cr
&=& {4 R_{\pi}\; e^{-i(l+n+1)\varphi_k +i{\pi\over 2}|l-n|}\;
\sin\pi\delta\over
\sqrt{k_{\perp}^4-2k_{\perp}^2(p_{\perp}^2+q_{\perp}^2)
+(p_{\perp}^2-q_{\perp}^2)^2}}\;a^{|l|}\;b^{|n|} \nonumber\\
&& \times\left[\Theta(l\geq 0)\Theta(n<0)(ab)^{-\delta} \left({1\over
b}\sqrt{E_p+sM}\sqrt{E_q-rM} - a \sqrt{E_p-sM}\sqrt{E_q+rM}
\right)\right. \nonumber\\
&& - \left. \Theta(l<0)\Theta(n\geq 0) (ab)^{\delta} \left(b
\sqrt{E_p+sM}\sqrt{E_q-rM} - {1\over a}\sqrt{E_p-sM}\sqrt{E_q+rM}
\right)\right].
\end{eqnarray}

The partial wave analysis of the pair production process caused by a photon
which passes the AB string shows a rather unexpected feature:
The process turns out to be forbidden unless the quantum numbers $l$ and $n$ 
of the outgoing electron and positron have opposite signs. This in turn
implies that (the expectation values of) their kinetic angular momentum 
projections, $[\vec{r} \times (\vec{p}-e\vec{A})]_3 = - i \partial_\varphi
-\phi$ have opposite signs. (For a detailed discussion see \cite{Audretsch95}.)
In the framework of a semiclassical picture this means that created charged
particles need to pass the magnetic string in opposite directions.
Apparently this is necessary for the ingoing photon to give the excess of
its radial momentum, $k_\perp - (p_\perp + q_\perp),$ to the string and to
create a real electron-positron pair from the vacuum.

We draw attention to another characteristic trait of the pair creation process.
It takes place although the incoming photon is not influenced by the magnetic 
string {\em directly}. The process happens since the created charged particles 
interact with the AB potential. The magnetic string distorts the states of the 
virtual electron-positron pairs in the vacuum. The incoming photon interacts  
with these virtual pairs, and it can transform them into real pairs if the 
conditions for momentum transfer (and energy conservation) are fulfilled.

\subsection{Differential cross section for scattering states}

It is now easy to calculate the matrix element for the electron-positron pair 
production by a photon with respect to the electron (\ref{epw}) and positron 
(\ref{ppw}) scattering states. For an incoming photon with momentum $\vec{k}$ 
and polarization $\lambda$ which creates electron and positron with momenta 
$\vec{p},\;\vec{q}$ and spins $s,\;r$, correspondingly, the matrix element of 
the process reads 
\begin{eqnarray} \label{pwme1}
M_{\lambda} &:=& -i \langle(\vec{q}, r), (\vec{p}, s)|S^{(1)}
|(\vec{k}, \lambda)\rangle \nonumber\\
&=& \sum_{l, n} c_l^{(e)\ast} c_n^{(p)}\;
\widetilde{M}_{\lambda}(j_p, j_q)\nonumber\\
&=& {e\sqrt{2}\sin\pi\delta\over{\sqrt{\omega_{k}E_{q}E_{p}}}}\;
{\delta(E_p+E_q-\omega_{k})\; \delta(p_3+q_3) \over
\sqrt{k_{\perp}^4-2k_{\perp}^2(p_{\perp}^2+q_{\perp}^2)
+(p_{\perp}^2-q_{\perp}^2)^2}}\;R_{\lambda} \;\Sigma_{\lambda}
\end{eqnarray}
where the coefficients $c_l^{(e)\ast}$ and $c_n^{(p)}$ are given by eq. 
(\ref{cc}) and we denote
\begin{eqnarray} \label{Ssp}
i\Sigma_{\sigma} &:=&  
\sum_{l, n} e^{il\varphi_{pk}+i(n+1)\varphi_{qk}} \;
a^{|l|}\;b^{|n|} \nonumber \\
&\times&\left[(ab)^{-\delta}\Theta(l\geq 0)\Theta(n<0)
\left(\sqrt{E_p+sM}\sqrt{E_q+rM} - {a\over
b}\sqrt{E_p-sM}\sqrt{E_q-rM} \right)\right. \nonumber\\
&-& \left. (ab)^{\delta} \Theta(l<0)\Theta(n\geq 0) 
\left(\sqrt{E_p+sM}\sqrt{E_q+rM} - {b\over
a}\sqrt{E_p-sM}\sqrt{E_q-rM} \right)\right], 
\end{eqnarray}
\begin{eqnarray}
\Sigma_{\pi} &:=& 
\sum_{l, n} e^{il\varphi_{pk}+i(n+1)\varphi_{qk}} \;
a^{|l|}\;b^{|n|}\nonumber \\
&\times&\left[ (ab)^{-\delta}[\Theta(l\geq 0)\Theta(n<0) \left({1\over
b}\sqrt{E_p+sM}\sqrt{E_q-rM} - a \sqrt{E_p-sM}\sqrt{E_q+rM}
\right)\right. \nonumber\\
&-& \left.  (ab)^{\delta} \Theta(l<0)\Theta(n\geq 0)\left(b
\sqrt{E_p+sM}\sqrt{E_q-rM} - {1\over a}\sqrt{E_p-sM}\sqrt{E_q+rM}
\right)\right].
\end{eqnarray}
with $\varphi_{pk} := \varphi_{p}-\varphi_{k}$, $\varphi_{qk} 
:= \varphi_{q}-\varphi_{k}$.

Performing the sums over $l,\,n$ we obtain for the polarization state 
$\lambda=\sigma$
\begin{eqnarray} \label{Ss}
i\Sigma_{\sigma} &=& (ab)^{-\delta}\left( b\sqrt{E_p+sM}\sqrt{E_q+rM}
- a\sqrt{E_p-sM}\sqrt{E_q-rM}\right) \Sigma \nonumber\\
&-& (ab)^{\delta}\left( a\sqrt{E_p+sM}\sqrt{E_q+rM} - b\sqrt{E_p-sM}
\sqrt{E_q-rM}\right)\;e^{-i\varphi_{pq}}\;\Sigma^{\ast}
\end{eqnarray}
and for the polarization state $\lambda=\pi$
\begin{eqnarray} \label{Sp}
\Sigma_{\pi} &=& (ab)^{-\delta}\left( \sqrt{E_p+sM}\sqrt{E_q-rM} -  ab
\sqrt{E_p-sM}\sqrt{E_q+rM}\right) \Sigma \nonumber\\
&-& (ab)^{\delta}\left( ab\sqrt{E_p+sM}\sqrt{E_q-rM} - \sqrt{E_p-sM}
\sqrt{E_q+rM}\right)\;e^{-i\varphi_{pq}}\;\Sigma^{\ast}
\end{eqnarray}
with
\begin{equation} \label{S}
\Sigma := {1\over 1-a\;e^{i\varphi_{pk}}}\cdot {1\over
1-b\;e^{-i\varphi_{qk}}}.
\end{equation}

Eqs. (\ref{pwme1}) together with (\ref{Ssp})--(\ref{S}) contain results for 
the pair production matrix elements with respect to scattering states.

\medskip

Based on the matrix element (\ref{pwme1}) we evaluate the differential 
probability of electron-positron pair production by a single photon per 
unit length of the magnetic string and unit time
\begin{equation} \label{dp}
dW_{\lambda}=W_{\lambda}\;p_{\perp}dp_{\perp} d\varphi_p
\;q_{\perp}dq_{\perp}d\varphi_q dp_3 dq_3
\end{equation}
where
\begin{equation} \label{W1}
W_{\lambda} := 
{|M_\lambda|^2\over T L}
{e^2\;\sin^2\pi\delta\over 8\pi^4 \omega_{k}E_{q}E_{p}}
\cdot {\delta(E_p+E_q-\omega_{k})\; \delta(p_3+q_3) \over
k_{\perp}^4-2k_{\perp}^2(p_{\perp}^2+q_{\perp}^2)
+(p_{\perp}^2-q_{\perp}^2)^2}\;R^2_{\lambda}\;|\Sigma_{\lambda}|^2.
\end{equation}
Calculating
\begin{equation} \label{R}
R_{\sigma}^2 = {1+s_3 r_3 \over 2}+{q_3^2\over q_3^2+M^2}{1-s_3 r_3 \over 2}, 
\quad R_{\pi}^2 = {1-s_3 r_3 \over 2}+{q_3^2\over q_3^2+M^2}{1+s_3 r_3 \over 2}
\end{equation}
where $s_3={\rm sign}\; s, \; r_3={\rm sign}\; r$ and sorting terms with 
respect to the flux parameter $\delta$
\begin{equation} \label{Sl}
|\Sigma_{\lambda}|^2 = (a b)^{-2\delta} P^{(-)}_{\lambda} + (a
b)^{2\delta} P^{(+)}_{\lambda} + P^{(0)}_{\lambda}
\end{equation}
we find
\begin{eqnarray} \label{P}
P^{(-)}_{\sigma} &=& [b^2 (E_p+sM)(E_q+rM) + a^2 (E_p-sM)(E_q-rM) - 2ab
p_{\perp} q_{\perp}]\;|\Sigma|^2, \nonumber \\
P^{(+)}_{\sigma} &=& [a^2 (E_p+sM)(E_q+rM) + b^2 (E_p-sM)(E_q-rM) - 2ab
p_{\perp} q_{\perp}]\;|\Sigma|^2, \nonumber \\
P^{(0)}_{\sigma} &=& [-2ab (E_p E_q +sr M^2) 
+ (a^2+b^2) p_{\perp}q_{\perp}]\;|\Sigma|^2\;F(\varphi),  \nonumber \\
P^{(-)}_{\pi} &=& [(E_p+sM)(E_q-rM) + a^2 b^2 (E_p-sM)(E_q+rM) - 2ab
p_{\perp} q_{\perp}]\;|\Sigma|^2, \nonumber \\
P^{(+)}_{\pi} &=& [a^2b^2 (E_p+sM)(E_q-rM) + (E_p-sM)(E_q+rM) - 2ab
p_{\perp} q_{\perp}]\;|\Sigma|^2, \nonumber \\
P^{(0)}_{\pi} &=& [-2ab (E_p E_q -sr M^2) 
+ (1+a^2b^2) p_{\perp}q_{\perp}]\;|\Sigma|^2\;F(\varphi)
\end{eqnarray}
where 
\begin{equation} \label{SS}
|\Sigma|^2 = {1\over (1+a^2-2a\cos\varphi_{pk})(1+b^2-2b\cos\varphi_{qk})},
\end{equation}
\begin{equation} \label{SF}
|\Sigma|^2 F(\varphi) := e^{i\varphi_{pq}}\Sigma^2 
+ e^{-i\varphi_{pq}}{\Sigma^{\ast}}^2
\end{equation}
with
\begin{equation} \label{F}
F(\varphi) :=
{[2a-(1+a^2)\cos\varphi_{pk}][2b-(1+b^2)\cos\varphi_{qk}] -
(1-a^2)(1-b^2)\sin\varphi_{pk}\sin\varphi_{qk}\over
(1+a^2-2a\cos\varphi_{pk}) (1+b^2-2b\cos\varphi_{qk})}.
\end{equation}

The differential probability of the pair production process with respect to 
the variables $\vec{p},\;\vec{q}$ is given by Eqs. (\ref{dp})--(\ref{F}). 
 
The complete information about energy, angular and polarization distributions 
of created electrons and positrons is contained in the effective differential 
cross section,
\begin{equation} \label{dcs1}
{d\sigma_{\lambda}\over d E_q d\varphi_q d\varphi_p dq_3}  =
{e^2\;\sin^2\pi\delta\over 32\pi^4
}{R^2_{\lambda}\;|\Sigma_{\lambda}|^2 \over \omega_k^3 (q_3^2 +
M^2)} \,.
\end{equation}
These distributions can be observed in experiments.
Instead of $p_{\perp}$ and $q_{\perp}$ we made use of more convenient 
variables -- the positron (electron) energy $E_q \; (E_p)$ and the 
$z$-component of the positron (electron) momentum $q_3 \;
(-p_3)$.  They are, obviously, related by the equalities 
$E_p+E_q=\omega_k, \; p_3+q_3=0.$ 

We rewrite the cross section (\ref{dcs1}) in a more detailed form
\begin{equation} \label{dcs2}
{d\sigma_{\lambda}\over d E_q d\varphi_q d\varphi_p dq_3}  =
{e^2\;\sin^2\pi\delta\over 128\pi^4 }{c^{-\delta}
S^{(-)}_{\lambda}+ c^{\delta}
S^{(+)}_{\lambda}+S^{(0)}_{\lambda}\over \omega^3_k (q_3^2 + M^2)
(E_p-p_{\perp}\cos\varphi_{pk}) (E_q -
q_{\perp}\cos\varphi_{qk})}
\end{equation}
with
\begin{equation} \label{c}
c := (ab)^2 = {(E_p-\sqrt{q_3^2+M^2})(E_q-\sqrt{q_3^2+M^2}) \over 
(E_p+\sqrt{q_3^2+M^2})(E_q+\sqrt{q_3^2+M^2})}
\end{equation}
and $p_\perp = \sqrt{E_p^2-q_3^2-M^2},\;q_\perp = \sqrt{E_q^2-q_3^2-M^2}.$

Then we have for the polarization state $\sigma$
\begin{eqnarray} \label{S1}
S^{(\mp)}_{\sigma} &=& (1+s_3r_3)(q^2_3+M^2)
\left[E^2_p + E^2_q -2(q^2_3+M^2) \mp s_3 (q^2_3+M^2)(E^2_p - E^2_q)\right] \\
&&+ (1-s_3r_3)  q_3^2 (E_p -  E_q)^2(1 \pm s_3), \nonumber\\
S^{(0)}_{\sigma} &=& - (1+s_3r_3) 2p_{\perp}q_{\perp} (q_3^2+M^2) F(\varphi)
\end{eqnarray}
and for the polarization state $\pi$
\begin{eqnarray} 
S^{(\mp)}_{\pi} &=& (1-s_3r_3)(q^2_3+M^2)\omega^2_k (1 \pm s_3 ) \nonumber\\
&&+ (1+s_3r_3) q_3^2 
\left[E^2_p+E^2_q-2(q^2_3+M^2) \mp s_3 (E_p^2-E_q^2) \right], \\
\label{S2}
S^{0}_{\pi} &=& (1+s_3r_3) 2p_{\perp}q_{\perp} q^2_3 F(\varphi)\,,
\end{eqnarray}
with $s_3 := {\rm sign}\,s$ and $ r_3 := {\rm sign}\,r$.

Eq. (\ref{dcs2}) together with (\ref{c})--(\ref{S2}) and (\ref{F}) gives 
the {\em final expression} for the effective differential cross section for 
the pair production process. We will discuss it at different energies of the 
incoming photon.

\section{The cross section for pair production at
different photon energies}

In this section we will analyze the angular and polarization distributions of 
the created electrons and positrons and find the total pair production cross 
section at different energies of the incoming photon. 

\subsection{Angular and polarization distributions}

The angular distributions for created electrons and positrons are of a fairly 
complicated nature. They simplify considerably at low and high photon 
energies. 

At {\em low photon energy}, just above the pair production threshold $2M,\;
\omega-2M \ll 2M,$ we have
$$
p_\perp \sim q_\perp \sim q_3 \ll M, \quad c\ll 1
$$
and
$$
S^{(\mp, 0)}_{\sigma} \sim S^{(0)}_{\pi} \ll S^{(\mp)}_{\pi} 
\approx 4M^4 (1-s_3r_3)(1\pm s_3).
$$
It means that electron-positron pair of low energies are created mainly from 
$\pi$-polarized photons, and the differential cross section for pair production 
above the threshold reads in this case 
\begin{equation} \label{dcsle}
{d\sigma_{\pi}\over d E_q d\varphi_q d\varphi_p dq_3} \approx
{e^2\;\sin^2\pi\delta\over 256\pi^4 } (1-s_3 r_3){c^{-\delta}
(1+s_3) + c^{\delta} (1-s_3)\over M^3}.
\end{equation}
The angular distributions for electrons and positrons of low energies are 
uniform in the plane perpendicular to the magnetic string but their dependence 
on the polar angle $\vartheta $ is rather intricate.

The polarizations of the electron and the positron depend strongly on the 
photon polarization state. 
Note that we are in the nonrelativistic limit and that $s_3$ ($r_3$) now
agrees with the spin projection of the electron (positron).
Created particles have spin projections of opposite 
signs, and electrons with positive spin projections (antiparallel to the 
magnetic string) are produced by $\pi$-polarized photons predominantly since 
$c\ll 1.$ Their fraction increases with the flux parameter $\delta.$
In this case the interaction of the magnetic moments with the string magnetic 
field is attractive for both the electron and the positron, and their wave 
functions are localized near the string.
The $\sigma$-polarized photon, on the other hand possesses a polarization 
vector perpendicular to the magnetic string and creates particles with spin 
projections $s_3$ and $r_3$ of equal signs which implies that their magnetic
moments have opposite directions. Due to the interaction of the magnetic 
moment with the magnetic field of the string therefore only one of the 
created particles, either the electron or the positron, is attracted to the 
string, which leads to an enhancement of the wave function near the string. 
This means that one of the particles is located near the string while 
the other one is located at a certain distance and, in a heuristic picture, 
it is unlikely that the $\sigma$-polarized photon creates an electron-positron 
pair.

With increasing photon energies the dependence on the photon polarization 
disappears since also larger orbital momenta contribute to the cross section.
This can be seen from eqs.~(\ref{ms2}) and (\ref{mp2}).

\medskip

At {\em high photon energies}, $\omega \gg M,$ the angular 
distributions are very 
simple. In this case the electron and positron are emitted predominantly in 
the forward direction, within a narrow cone surrounding the direction of 
motion of the photon. Due to the presence the factors 
$E_p-p_{\perp}\cos\varphi_{pk}$, 
$E_q-q_{\perp}\cos\varphi_{qk}$ and $q^2_3+M^2 \sim E^2_q \cos^2\vartheta$ 
in the denominator of (\ref{dcs2}) their angular distributions have sharp 
maxima in this direction 
($\varphi_{pk}\sim\varphi_{qk}\sim 0,\;\vartheta\sim \pi/2$) 
and the effective angular aperture of the cone is given in order of magnitude 
by $M /\omega_k.$

\subsection{The total cross section}

Let us now analyse the energy behavior of the total cross section. 
Integration  of the differential cross section (\ref{dcs2}) over the 
azimuthal angles
$\varphi_{p},\;\varphi_{q}$ leads to an additional factor 
$4\pi^2 /(q_3^2+M^2)$ and removes the term with $S^{(0)}_{\lambda}$ 
from the cross section,
\begin{equation} \label{dcs3}
{d\sigma_{\lambda}\over d E_q dq_3}  = {e^2\;\sin^2\pi\delta\over32\pi^2}\;
{c^{-\delta}S^{(-)}_{\lambda}+c^{\delta}S^{(+)}_{\lambda}\over
\omega^3_k (q_3^2 + M^2)^2}.
\end{equation}
Performing the sums over polarizations of created electron and positron
we obtain
\begin{equation} \label{dcs4}
{d\sigma_{\lambda}\over d E_q dq_3}  =
{e^2\;\sin^2\pi\delta\over 8\pi^2 }{c^{-\delta} 
+ c^{\delta}\over \omega^3_k (q_3^2 + M^2)^2}\;C_{\lambda}\end{equation}
with
\begin{eqnarray} \label{S3}
C_{\sigma} &:=& (q^2_3+M^2)[E^2_p+ E^2_q - 2(q^2_3+M^2)] 
+ q_3^2 (E_p- E_q)^2, \nonumber \\
C_{\pi} &:=& (q^2_3+M^2) \omega_k^2 + q_3^2 [E^2_p+ E^2_q - 2(q^2_3+M^2)].
\end{eqnarray}

Since the variables $p_\perp$ and $q_\perp$ are both positive the variable 
$q_3$ ranges from $-q_3^{\rm max}$ to $q_3^{\rm max}$ where 
$q_3^{\rm max}={\rm min}\left(\sqrt{E_q^2 - M^2}, \sqrt{E_p^2 - M^2}\right)$. 
Introducing new  variables 
\begin{equation}
\varepsilon = |E_p-E_q|\,, \quad
x=\sqrt{q_3^2+M^2} \,,
\end{equation}
we obtain the general expression for the total cross section for pair 
production by a photon of the energy $\omega_k$ and polarization $\lambda$ 
\begin{eqnarray} \label{tcs1}
\sigma_{\lambda} &=& {e^2\;\sin^2\pi\delta\over 4\pi^2}\;{1 \over\omega^3_k}\; 
\int_0^{\omega_k -2M}d\varepsilon \int_M^{\omega_k-\varepsilon\over 2} dx\; 
{c^{-\delta}+c^{\delta}\over 
x^3 \sqrt{x^2-M^2}}\;C_{\lambda}(\varepsilon, x) \nonumber\\
&=& {e^2\;\sin^2\pi\delta\over 4\pi^2}\;{1 \over\omega^3_k}\;
\int_M^{\omega_k \over 2} dx\; {1\over x^3 \sqrt{x^2-M^2}}\int_0^{\omega_k -2x} 
d\varepsilon \; (c^{-\delta}+c^{\delta})\;C_{\lambda}(\varepsilon, x) 
\end{eqnarray}
where 
\begin{eqnarray} \label{c,si.pi}
c &=& (ab)^2 = {(\omega_k-2x)^2-\varepsilon^2\over 
(\omega_k+2x)^2-\varepsilon^2}, \\
C_{\sigma}(\varepsilon, x) &=& {1\over 2} x^2(\omega_k^2+\varepsilon^2 -4x^2) 
+ (x^2-M^2)\varepsilon^2,\\
C_{\pi}(\varepsilon, x) &=& x^2 \omega_k^2+{1\over 2}(x^2-M^2) 
(\omega_k^2+\varepsilon^2-4x^2).
\end{eqnarray}

The remaining integrals over $\varepsilon$ and $x$ can not be found
analytically for arbitrary values of the flux parameter $\delta$. Even for the 
symmetric case $\delta={1\over 2}$ we have a rather complicated integral
\begin{equation} \label{tcs2}
\sigma_{\lambda} = {e^2\over 2\pi^2}\;{1 \over\omega^3_k}\;
\int_M^{\omega_k\over 2} dx\;{1\over x^3\sqrt{x^2-M^2}} 
\int_0^{\omega_k -2x} d\varepsilon  {(\omega_k^2-\varepsilon^2+4 x^2)\;
C_{\lambda}(\varepsilon, x)\over 
\sqrt{(\omega_k^2-\varepsilon^2)^2- 8x^2(\omega_k^2+x^2)+16x^4}}.
\end{equation}
But the general expression (\ref{tcs1}) for the total cross section of pair 
production simplifies considerably at low and high photon energies to which
we turn now. 

\medskip

At {\em low energies}, near the pair creation threshold, 
$\omega_k - 2M \ll M$ we have $C_{\sigma} \ll C_{\pi} \approx  4M^2$.
Therefore we will consider only the polarization state $\pi$.
It is then 
$$
c \approx {(\omega_k-2x)^2-\varepsilon^2\over 16 M^2} \ll 1. 
$$
Dropping in (\ref{tcs1}) the term $c^{\delta} \ll c^{-\delta}$ and introducing 
new dimensionless variables $t$ and $y$ by 
$$
\varepsilon = (\omega_k-2x) t, \quad x = M +{\omega_k-2M\over 2} y, $$
we obtain
\begin{eqnarray}\label{tcsle}
\sigma_{\pi} &\approx& {e^2\;\sin^2\pi\delta\over 8\pi^2 M^2}\; 
\int_M^{\omega_k\over 2} dx\;\int_0^{\omega_k-2x} d\varepsilon 
{c^{-\delta}\over \sqrt{2M (x-M^)}} \nonumber\\
&=& {e^2\;\sin^2\pi\delta\over 4\sqrt{2}\pi^2 M}\; 
\left({\omega_k-2M \over 2M}\right)^{{3\over 2}-2\delta} 
B(1-\delta, 1-\delta)\;B({1\over 2}, 2-2\delta) \nonumber\\
&=& {r_0\;\sin^2\pi\delta\over \sqrt{2}\pi}\;
\left({\omega_k-2M \over 2M}\right)^{{3\over 2}-2\delta} 
B(1-\delta, 1-\delta)\;B({1\over 2}, 2-2\delta)
\end{eqnarray}
where $r_0={e^2\over 4\pi M}$ is the classical electron radius and the constant 
$B(\mu, \nu)$ is the Euler's integral of the first kind. After integration over 
a small energy interval $\Delta$ above the threshold, 
$2M \ge \omega_k \le 2M (1+\Delta)$ the integral cross section for pair 
production reads
\begin{equation} \label{I}
I := \int_{2M}^{2M(1+\Delta)}\sigma_{\pi}(\omega_k)\,d\omega_k 
\approx {e^2\,\sin^2\pi\delta\over 2\sqrt{2}\pi}\;
{\Delta^{{5\over 2}-2\delta}\over {5\over 2}-2\delta}\;
B(1-\delta, 1-\delta)\;B({1\over 2}, 2-2\delta).
\end{equation}
This quantity determines the output of electron-positron pairs produced by 
a photon per unit length of the magnetic string per unit time within the given 
energy interval. At $\delta = {1\over 2}$ we find
\begin{equation}
I = {e^2\sqrt{2}\over 3\pi}\;\Delta^{3\over 2}.
\end{equation}

\medskip

At {\em high photon energies}, $\omega_k \gg M,$ the parameter $c$ behaves 
like $c \sim 1$. In this case we have
\begin{equation} \label{tcshe1}
\sigma_{\lambda} \approx {e^2\;\sin^2\pi\delta\over 2\pi^2}\;
{1 \over\omega^3_k}\; \int_M^{\omega_k \over 2} dx \;
{1\over x^3 \sqrt{x^2-M^2}}\;\int_0^{\omega_k -2x} d\varepsilon \; 
C_{\lambda}(\varepsilon, x).
\end{equation}
Calculating the integral over $\varepsilon$ we find that the main contribution 
to the asymptotic behavior of the cross section at $\omega_k \gg M$ arises from 
values of $x\sim M.$ Performing  integration over $x$ we obtain 
\begin{equation} \label{tcshe2}
\sigma_{\lambda} 
\approx {e^2\,\sin^2\pi\delta\over 4\pi M}\, a_{\lambda}
= r_0\,\sin^2\pi\delta\, a_{\lambda} 
\end{equation}
with $a_{\sigma}={2\over 3}$ and $a_{\pi}=1.$

At high photon energies the total cross section of the pair production tends 
asymptotically to constant values for both photon polarizations. This energy 
dependence is compatible with unitarity. (One might have expected that the 
singular {\it pure AB potential} leads to an increasing cross section and 
causes the violation of the perturbative theory at high energies, in a similar
way as in the case of the idealized, infinitely thin cosmic string 
\cite{Skarzhinsky94}). 

However, we do not consider the high energy AB pair 
production as a realistic subject for experimental investigation. For a 
realization
one needs to take much care about coherence of the high energy photon beam. 
We are going to estimate the possibility of the experimental observation of 
the AB pair production effect in a subsequent paper.


\section{Conclusion}

We have analyzed the electron-positron pair production by a single photon 
under the influence of a magnetic string in first order perturbation theory, 
which, as other quantum processes connected with the AB effect, leads to 
rather unexpected results. Photons do not interact directly with magnetic 
fields, and the process which was considered here happens due to the 
interaction of the created charged particles in the final states with the 
AB potential.

In addition to the AB interaction, resulting from the non-integrable phase
factors, which all quantum particles suffer, spin particles interact with the
magnetic field via their magnetic moments. This strongly influences their
behavior near the flux tube. In the idealized case of an infinitely thin
magnetic string their wave functions do not vanish on the string and the
non-locality of the AB effect is modified by a local interaction. This
interaction leads to a specific behavior of the cross section.

We evaluated the differential cross section for the pair production, which 
contains complete information about energy, angular and polarization 
distribution of the created particles, as well as the total cross section 
and analysed them for different energy regimes. For low photon energies, 
just above the pair production threshold, electrons and positrons are produced 
predominantly by the $\pi$-polarized photons with polarization vectors 
directed along the magnetic string. This result may be the most interesting 
one with regard to possible experimental observations of the AB pair 
production process. 

Of course, the observation of the AB effect, which is done by means of 
electron interference and electron holography \cite{Peshkin89}, is not a 
simple task, and the experiments with photons which interact with the magnetic 
string and create electron-positron pairs requires a careful discussion. For 
these photons of rather high energies there exist additional effects which 
can obscure the AB pair production. In particular, this is the pair 
production by the photon in collision with material of the tube carrying 
the magnetic flux.
 
We draw attention to a remarkable feature which is characteristic for 
quantum processes in the presence of the AB string both for spinless and 
for spin particles. The pair production process happens if the created 
electrons and positrons have angular momentum projection of opposite signs. 
In a sense the virtual charged particles need to circle the AB string to 
transform to real ones. In this case the photon can transmit a part of its 
perpendicular momentum to the string. We analyzed in detail how the total 
cross section of the process depends on the photon polarization. This 
analysis may be very important because it enables to distinguish the pure 
effect from interfering effects accompanying the AB pair production process. 

Finally we point out the analogy to the pair production process in the 
presence of a cosmic string \cite{Skarzhinsky94}. In this case an additional 
term appears in the Dirac equation which results from the spin connection. It 
corresponds to the vector potential term in the AB case.


\section*{Acknowledgments}

V.~S.~thanks J.~Audretsch and the members of his group at
the University of Konstanz for hospitality, collaboration and many
fruitful discussions. This work was supported by the Deutsche
Forschungsgemeinschaft


\end{document}